\def\apj{ApJ}
\def\apjl{ApJL}
\def\mnras{MNRAS}

\def\aj{AJ}

\def\apjs{ApJS}

\newcommand{\x}{$\times$}
\newcommand{\about}{$\sim$}
\newcommand{\hi}{{H{\sc i}}}
\newcommand{\Mstar}{M$_*$}
\newcommand{\Msun}{M$_\odot$}

\newcommand{\fgas}{$f_{\rm gas}$}

\documentclass[
    ,final            % use final for the camera ready runs
  ]
  {aipproc}

\layoutstyle{6x9}

\begin{document}

\title{The GALEX Arecibo SDSS Survey (GASS)}

%95.80.+p 	Astronomical catalogs, atlases, sky surveys, databases, retrieval systems, archives, etc.
%95.85.-e 	Astronomical observations (additional primary
%    heading(s) must be chosen with these entries to represent the astronomical objects and/or properties studied)
%95.85.Bh 	Radio, microwave (>1 mm)
%95.85.Kr 	Visible (390-750 nm)
%95.85.Mt 	Ultraviolet (10-300 nm)
%98.62.Ve 	Statistical and correlative studies of properties (lum. and mass functions; M/L ratio; TFR, etc.)

\author{Barbara Catinella}{
  address={Max-Planck-Institut f\"{u}r Astrophysik, D-85741 Garching, Germany}
}

\author{David Schiminovich}{
  address={Department of Astronomy, Columbia University, New York, NY 10027, USA}
}

\author{Guinevere Kauffmann}{
  address={Max-Planck-Institut f\"{u}r Astrophysik, D-85741 Garching, Germany}
}

\begin{abstract}
The GALEX Arecibo SDSS Survey (GASS) is a large targeted survey that
started at Arecibo in March 2008. GASS is designed to measure the neutral
hydrogen content of \about 1000 massive galaxies (with stellar
mass \Mstar >$10^{10}$ \Msun) at redshift $0.025 < z < 0.05$, uniformly
selected from the SDSS spectroscopic and GALEX imaging surveys.
Our selected mass range straddles the 
recently identified ``transition mass'' (\Mstar \about 3 \x $10^{10}$ \Msun)
above which galaxies show a marked decrease 
in their present to past-averaged star formation rates.
GASS will produce the first statistically significant
sample of massive ``transition'' galaxies with homogeneously measured
stellar masses, star formation rates and gas properties. 
The analysis of this sample will allow us
to investigate if and how the cold gas responds to a variety of
different physical conditions in the galaxy, thus yielding insights on
the physical processes responsible for the transition between blue,
star-forming and red, passively evolving galaxies. GASS will be of
considerably legacy value not only in isolation but also by
complementing ongoing \hi -selected surveys.
\end{abstract}

\maketitle

\section{Introduction}

While the clear distinction between red and old ellipticals
and bluer and star-forming spirals has been known for a long
time, recent work has shown that galaxies appear 
to divide into two distinct ``families'' at a stellar mass 
\Mstar \about 3 \x $10^{10}$ \Msun\ \cite{str01,kau03,bal04}.
Lower mass galaxies typically have young stellar populations, low
surface mass densities and the low concentrations characteristic of
disks. On the other hand, galaxies with old stellar populations, high
surface mass densities and the high concentrations typical of bulges
tend to have higher mass, \Mstar > 3 \x $10^{10}$ \Msun.
It is clearly important to understand why there should be a
characteristic mass scale where galaxies transition 
from young to old. And, in order to understand how such transition
takes place, it is critical to study the cold \hi\ gas, which is the 
source of the material that will eventually 
form stars. 

\hi\ studies of transitional galaxies are currently not possible using
existing \hi\ surveys, which sample only shallow volumes: a
specifically designed, targeted survey is required. The GALEX Arecibo
SDSS Survey (GASS; P.I.: D. Schiminovich) is such a survey, designed
to measure the neutral hydrogen content of a representative sample of
massive, transitional galaxies, uniformly selected from the
Sloan Digital Sky Survey (SDSS, \cite{sdss}) spectroscopic data base
and {\em Galaxy Evolution Explorer} (GALEX, \cite{galex}) imaging surveys (Fig. 1). The
final data base will include
optical, UV and \hi\ parameters for \about 1000 galaxies with stellar mass
\Mstar $>10^{10}$ \Msun\ and gas mass fractions as low as 1.5\%. 
Several key questions will be addressed using these data: \\
$-$ What fraction of massive galaxies possess reservoirs of cold gas
but no active star formation (SF)? Is this population distinguished by its
environment, presence of an active galactic nucleus (AGN) and/or
morphology? What implications does this have for the gas accretion
history of the galaxies and SF thresholds? \\
$-$ Which populations are gas deficient (at fixed \Mstar\ and/or
fixed SF rate) with respect to the complete 
sample? Can we identify a quenching process, starvation or gas removal
process that can explain this deficiency? \\
We seek to answer these questions by performing an unbiased survey,
targeting galaxies in the SDSS spectroscopic sample selected only by
their redshift and stellar mass.

\begin{figure}
\centering
  \includegraphics[height=0.46\textheight]{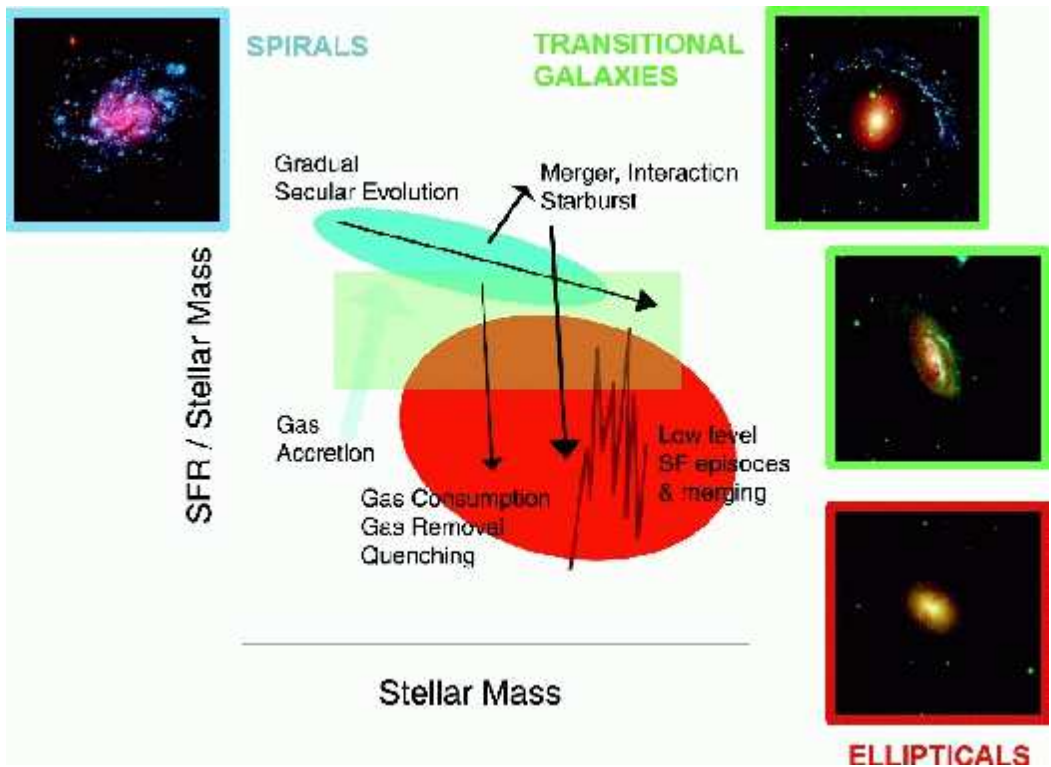}
  \caption{Schematic view of star formation history vs. stellar mass,
  adapted from \cite{sch07}. GASS will shed light on the physical processes responsible
  for the transition between blue, star-forming galaxies and red sequence.}
\end{figure}

\section{Sample selection}

The selection criteria for the GASS targets are the following: \\
(1) Location within the intersection of the footprints of the SDSS DR6
primary spectroscopic survey, the GALEX Medium Imaging Survey (MIS)
and the Arecibo Legacy Fast ALFA (ALFALFA; \cite{alfalfa}) survey.
ALFALFA is an ongoing, blind \hi\ survey of the extragalactic sky visible
from Arecibo. Existing ALFALFA coverage greatly increases our survey
efficiency $-$ the gas-richest GASS targets (roughly 20\% of the
sample) will be already detected by the shallow ALFALFA observations.\\
(2) Redshift $0.025 < z < 0.05$\\
(3) Stellar mass $\rm 10.0 < log ~(M_*/M_{\odot}) < 11.5$.\\ 
The targets will be observed until either a gas fraction limit 
$f_{\rm gas} \equiv {\rm M_{HI} / M_*} =1.5\%$
or a minimum \hi\ mass of $10^{8.5}$ \Msun\ is reached. 

As Fig. 2 shows, a complete sample of galaxies with \Mstar $>10^{10}$
\Msun\ spans a very wide range in NUV/optical colors. 
The spread is much more dramatic than that seen in optical colors
alone. Galaxies with NUV$-r < 3$ are typically star-forming spirals
and are likely to have reasonably high gas fractions. Passive ellipticals
have redder colors, typically NUV$-r > 5$. The intermediate color regime,
where the transitional galaxies are found, is consistent with ongoing
SF rate at a level of a few percent of \Mstar\ over the past
Gyr and moderate gas fractions (1$-$10\%). Over this \Mstar\
regime, ALFALFA detects only systems with \fgas > 10\%. 
Thus it is clear that, in order to systematically quantify the \hi\
properties of these objects, a significantly deeper, targeted \hi\
survey is required.

\begin{figure}
\centering
  \includegraphics[height=0.195\textheight]{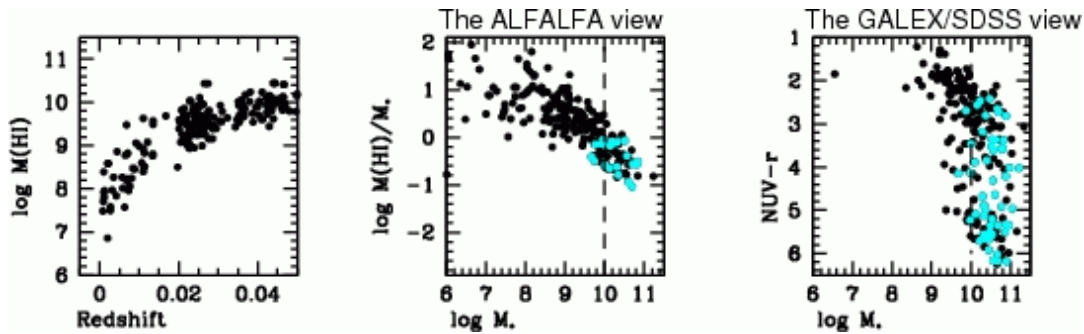}
  \caption{Left: \hi\ mass vs. redshift for a sample of 227 galaxies from
  ALFALFA with matches in the SDSS spectroscopic
  sample. Center: \hi\ gas mass fraction vs. stellar mass. Right:
  Properties of a random sample of 227 galaxies from GALEX/MIS with
  matches in SDSS and $0.01<z<0.05$. Cyan points denote AGN. Dashed
  line: GASS stellar mass limit.} 
\end{figure}

\section{GASS data}

GASS will deliver catalogs of \hi\ and value-added 
properties which will be used as the basis for a 
large number of studies. The \hi\ spectral data pro\-ducts will be
incorporated into the Cornell-NAIC Extragalactic HI Digital Archive
({\em http://arecibo.tc.cornell.edu/hiarchive}), a 
registered US National Virtual Observatory node that contains the
ALFALFA data releases as well as other \hi\ data sets. 
More details, including the current status of the survey and the list of team
members, can be found on the GASS web site
({\em http://www.mpa-garching.mpg.de/GASS}).

%\begin{theacknowledgments}
%\end{theacknowledgments}

\bibliographystyle{aipproc}   % if natbib is available

%\bibliography{biblio}

\end{document}